\documentclass[aps,prd,twocolumn,amsmath,superscriptaddress,amssymb,showpacs,floatfix,nofootinbib]{revtex4-1}

\usepackage{hyperref}
\usepackage{graphicx}
\usepackage{dcolumn}
\usepackage{xcolor}
\usepackage{bm}
\usepackage{natbib}
\usepackage{calligra}
\usepackage[T1]{fontenc}
\usepackage{egothic}
\usepackage[T1]{fontenc}
\newfont{\rsfsten}{rsfs10 scaled 1200}
\newfont{\rsfsseven}{rsfs10 scaled 1200}
\newfont{\rsfsfive}{rsfs10 scaled 1200}
\usepackage{epsfig}
\usepackage{units}
\usepackage[utf8]{inputenc}
\usepackage{yfonts}

\newcommand{\be}{\begin{equation}}
\newcommand{\ee}{\end{equation}}
\newcommand{\bea}{\begin{eqnarray}}
\newcommand{\eea}{\end{eqnarray}}

\newcommand{\Mpc}{{\rm ~Mpc}}

\newcommand{\Gpc}{{\rm ~Gpc}}

\newcommand{\yr}{{\rm ~yr}}
\newcommand{\tzo}{{\rm T\dot{\rm Z} \rm O}}

\newcommand{\BHprime}{BH$^{\prime} \,$}

\begin{document}

\title{Can Thorne-$\dot{\textrm{Z}}$ytkow Objects source GW190814-type events?}

\author{Ilias Cholis}
\email{cholis@oakland.edu, ORCID: orcid.org/0000-0002-3805-6478}
\affiliation{Department of Physics, Oakland University, Rochester, Michigan, 48309, USA}
\author{Konstantinos Kritos}
\email{ge16004@central.ntua.gr}
\affiliation{Physics Division, National Technical University of Athens, Zografou, Athens, 15780, Greece}
\author{David Garfinkle}
\email{garfinkl@oakland.edu, ORCID: orcid.org/0000-0003-4415-774X}
\affiliation{Department of Physics, Oakland University, Rochester, Michigan, 48309, USA}

\date{\today}

\begin{abstract}
        The LIGO-Virgo collaboration reported in their third run the coalescence event GW190814
        involving a 2.6 $M_{\odot}$ object with a 23 $M_{\odot}$ black hole. In this letter we study the 
        conditions under which Thorne-$\dot{\textrm{Z}}$ytkow objects ($\tzo$s) can be connected to that 
        type of events. 
	We evaluate first the rate of appearance of $\tzo$s in the local Universe. Under the assumption 
	that $\tzo$s eventually become low mass gap black holes we evaluate how those black holes end up in binaries 
	with other stellar mass black holes and compare to the reported rate for GW190814-type 
	of events (1-23 $\Gpc^{-3} \yr^{-1}$). We find that $\tzo$s in dense stellar clusters can not
	explain the LIGO-Virgo rate without a $\tzo$ population in the field providing a dominant contribution. 
	We also find that $\tzo$s formed within hierarchical triple systems in the field with the third more
	distant star being the progenitor of a stellar mass black hole, may be able to give a rate comparable
	to that of GW190814-type events. In that case, future observations should discover mergers
	between stellar mass and low mass gap black holes, with the lower mass spanning the entire low 
	mass gap range.
	
\end{abstract}

\maketitle

\textit{Introduction:}
The recent detection of the LIGO-Virgo coalescence event GW190814, where a  
2.6 $M_{\odot}$ object merged with a 23 $M_{\odot}$ black hole represents a new class 
of gravitational wave (GW) events, involving compact objects with mass in the low mass gap 
range of 2.5-5 $M_{\odot}$.  The 2.6 $M_{\odot}$ object may be the lightest black hole (BH), 
or the most massive neutron star (NS) ever observed \cite{Ozel:2016oaf, Fishbach:2017zga, 
Wysocki:2018mpo, Thompson:2018ycv, Takhistov:2020vxs, Tews:2020ylw,  Tan:2020ics, 
Zhang:2020zsc, Most:2020bba}. Such a detection may suggest that objects with such 
masses are more commonly found than previously suggested. Potential pathways to objects with 
masses in the low mass gap range may involve objects that are formed from NSs by accretion of 
matter (see e.g. \cite{Safarzadeh:2020ntc}). One of these pathways can be Thorne-$\dot{\textrm{Z}}$ytkow 
objects ($\tzo$s) that contain a NS inside a red giant \cite{1975ApJ...199L..19T, 1977ApJ...212..832T}. 
The final object may be a BH in the low mass gap range. Given that $\tzo$s require tight binary systems 
as their starting point we expect that such objects are mostly found in regions rich in stars. 

In this \textit{letter} we examine the conditions under which  low mass gap BHs sourced from $\tzo$s 
can form binaries with regular stellar mass range BHs (i.e. $\ge 5 M_{\odot}$) that are tight enough to 
merge within a Hubble time and at a rate similar to that of the GW190814 class, evaluated 
by the LIGO-Virgo collaboration to be $1-23 \Gpc^{-3}\yr^{-1}$ \cite{Abbott:2020khf}. We find 
that while dense stellar environments can not contribute significantly to the observed rate, hierarchical 
triple systems in the field may be able to explain events as the GW190814. 

\textit{An upper bound estimation of  $\, \tzo$s becoming low mass gap BHs:}
Using massive X-ray binaries  Ref. \cite{1995MNRAS.274..485P} estimated that in a galaxy of the 
size of the Milky Way, the emergence rate of $\tzo$s is $1-2 \times 10^{-4} \yr^{-1}$. More recent estimates 
suggest a rate of at least $0.9\times 10^{-4} \yr^{-1}$ \cite{Michaely:2016krw} and $\simeq 1.5\times 10^{-4} 
\yr^{-1}$ \cite{2018JApA...39...21H}. We will use from here on that $\tzo$s with a massive red giant, 
emerge with a rate of $1.5 \times 10^{-4} \yr^{-1}$ per Milky Way-like galaxy \footnote{A NS forming a 
binary with a red giant with a semi-major axis of $\sim 1$ AU or less can merge with its He core while 
it is in the common envelope phase \cite{Fryer:1999qs}. That pathway to the low mass gap BH is similar 
to that of the $\tzo$s and would only enhance the production rate of low mass gap BHs.}.  We will focus 
on the $\tzo$s where the red giant's initial mass was at least 5 $M_{\odot}$ and refer to them from now 
on as simply $\tzo$s. The number of $\tzo$s in a galaxy is directly proportional to the number of its massive 
binaries and thus also roughly proportional to the number of its stars, and as a result to a galaxy's total 
stellar mass $M^{\star}$. Normalizing to the rate of the Milky Way that has  $M^{\star}_{MW} = 10^{10.79} 
\, M_{\odot}$, a galaxy of stellar mass $M^{\star}$ will have a $\tzo$ emergence rate of,
\begin{equation}
\textrm{\textfrak{R}}_{\rm gal}^{\tzo}(M^{\star}) = 1.5 \times 10^{-4} \left(\frac{M^{\star}}{10^{10.79} 
\, M_{\odot}}\right) \yr^{-1}.
\label{eq:PerGalaxyRate}
\end{equation}
 
The Sloan Digital Sky Survey (SDSS) has measured the number density of galaxies in the redshift 
range of 0.02-0.06 \cite{2016MNRAS.459.2150W}. The number density of galaxies in a mass bin 
$d M^{\star}$ follows a Schechter mass function $\Phi$,
\begin{eqnarray}
n_{\rm gal} &=& \Phi(M^{\star}) d M^{\star} \\
                  &=& \Phi^{\star} e^{-M^{\star}/M^{\star}_{0}} \left( \frac{M^{\star}}{M^{\star}_{0}}\right)^{\alpha} 
                  d M^{\star}. \nonumber
\label{eq:galaxy_density}
\end{eqnarray}
$\Phi^{\star}$ is the overall normalization, $\alpha$ is the mass function power-law and $M^{\star}_{0}$ sets
an exponential suppression at at high masses. Late and early type galaxies have separate mass functions. 
Their combined mass function is (expressed in log$M^{\star}$) \cite{2016MNRAS.459.2150W},
\begin{eqnarray}
\Phi d \textrm{log}_{10} M^{\star} &=& ln(10) \, e^{-M^{\star}/M^{\star}_{0}} \, \Big  \{  \Phi^{\star}_{1} 
\left( \frac{M^{\star}}{M^{\star}_{0}}\right)^{\alpha_{1}+1} \nonumber \\
&+&  \Phi^{\star}_{2} \left( \frac{M^{\star}}{M^{\star}_{0}}\right)^{\alpha_{2}+1} \Big  \} 
d \textrm{log}_{10} M^{\star}. 
\label{eq:DoubleSchechtMF}
\end{eqnarray}
The appropriate normalizations are $\Phi^{\star}_{1} = h^{3} 10^{-3.31}$ $\Mpc^{-3}$, 
$\Phi^{\star}_{2} = h^{3} 10^{-2.01}$ $\Mpc^{-3}$ with power-laws $\alpha_{1} = -1.69$ and 
$\alpha_{2}=-0.79$ for $M^{\star}_{0} = 10^{10.79} M_{\odot}$ and $h = 0.7$ 
\cite{2016MNRAS.459.2150W}. 

Eqs.~\ref{eq:PerGalaxyRate}-\ref{eq:DoubleSchechtMF} give a local ($z \ge 0.06$) $\tzo$ 
emergence rate density of,
\begin{equation}
R^{\tzo}(z<0.1) = \int_{M^{\star}_{\rm min}}^{M^{\star}_{\rm max}} d M^{\star} \; 
\textrm{\textfrak{R}}_{\rm gal}^{\tzo}(M^{\star}) \frac{\Phi(M^{\star})}{M^{\star}}.
\label{eq:TZORateDensity}
\end{equation}
Taking $M^{\star}_{\rm min} = 10^{9} M_{\odot}$ ($M^{\star}_{\rm min} = 10^{10} M_{\odot}$) 
and $M^{\star}_{\rm max} = 10^{11.5} M_{\odot}$ we get a local $\tzo$ emergence rate density of 
$1.2 \times 10^{3}$ ($1.0 \times 10^{3}$) $\Gpc^{-3} \yr^{-1}$. 

As all $\tzo$s composed by a NS and a massive red giant are going to end up in a BH, we can 
evaluate the rate by which BHs are created just from these objects \footnote{$\tzo$s where the 
original companion of the NS was a low mass star may still give a NS as a final product. All our rates 
here rely on the observational constraints where the companion will become a massive red giant.}. 
To distinguish these BHs from those that originated directly from regular 
core-collapse we will denote these as BH$^{\tzo}$. Their rate is, 
\begin{equation}
    \mathcal{R}_{BH^{\tzo}}(z)=\int_0^z dz' R^{\tzo}(z')  \ {dV_c\over dz'}\ (1+z')^{-1},
\label{eq:observedRate}    
\end{equation}
where $dV_c/dz$ is the comoving volume element. 
Integrating to redshift of 0.1 we get $\mathcal{R}_{BH^{\tzo}}(0.1) 
= 3.8 \times 10^{2}$ $\yr^{-1}$ from galaxies with stellar mass of $M^{\star} \geq 10^{9} 
M_{\odot}$ and $\mathcal{R}_{BH^{\tzo}}(0.1) = 3.3 \times 10^{2}$ $\yr^{-1}$ from galaxies 
with stellar mass of $M^{\star} \geq 10^{10} M_{\odot}$; making our estimates insensitive 
to the low-mass end of the galaxies mass-function. 

Only the BH$^{\tzo}$ falling in the low mass gap are of importance here. These will be created if
the initial NS accretes at least 1 $M_{\odot}$ of mass~\cite{1993ApJ...411L..33C, 1995ApJ...440..270B}. 
Assuming that the neutron star of the $\tzo$ accreted 1/5th of the mass of the red giant, only stars with 
initial mass between 6 and 18 $M_{\odot}$ will give BH$^{\tzo}$ inside the low mass gap. Relying on the 
Kroupa initial mass function~\cite{2002Sci...295...82K}, with its uncertainties we find that 60-90$\%$ 
of the $\tzo$s will lead to a such a low mass BH$^{\tzo}$. Using the central values of \cite{2002Sci...295...82K} 
\footnote{Assuming a Kroupa initial mass function that scales with the mass of the star 
$m_{\rm star}$ as $\propto m_{\rm star}^{-2.3}$.} we get a local formation rate density of  BH$^{\tzo}$ 
with mass in the low mass gap of $9 \times 10^{2}$ $\Gpc^{-3} \yr^{-1}$. That rate is based on active 
star formation regions and may have been even larger in past epochs. Our rate is insensitive 
to the exact fraction of the mass of the red giant that is being absorbed. As an example, if the 
NS accretes 1/3rd of the total mass of the giant, the birth rate density varies by only $\simeq 5 \%$. 

Of the Milky Way's stellar mass about 1/3rd is in the bulge \cite{2016A&A...587L...6V, 
2016ARA&A..54..529B, 2018A&A...618A.147Z}. For small elliptical galaxies that are formed in a single 
epoch of collapse of gas as the much smaller in mass globular clusters have, the fraction of the stellar 
mass in dense regions may be even higher than for barred spiral galaxies (as the Milky Way). Also 
for the massive elliptical galaxies that are the result of mergers of smaller galaxies the relevant fraction will be 
as large as their progenitor galaxies.  As a result of the local birth rate density of BH$^{\tzo}$ with 
mass in the low mass gap that we estimated to be $9 \times 10^{2}$ $\Gpc^{-3} \yr^{-1}$, 
$3 \times 10^{2}$ $\Gpc^{-3} \yr^{-1}$ will be in dense stellar environments where the BH$^{\tzo}$ 
can interact and subsequently merge with other massive BHs.

\textit{Forming binaries of a low mass gap and a regular stellar mass BH:}
We consider two distinct paths for the formation of binaries composed of a BH$^{\tzo}$ within the 
low mass gap and a BH of 5 $M_{\odot}$ or larger. In the first one, the BH$^{\tzo}$ is initially in 
no binary and only after dynamical interactions forms a binary system with another BH. In the 
second pathway, originally there is a hierarchical triple containing a tight binary forming the $\tzo$ 
and a third object that will evolve to a stellar mass BH. After the death of the $\tzo$ and the 
formation of its BH$^{\tzo}$ since its natal kick is weak it will remain in a binary with the stellar mass BH.

\textit{Forming binaries inside globular clusters:}
In globular clusters all stars enter the main sequence at approximately the same moment. Thus 
the massive red giants of 6 to 18 $M_{\odot}$ are present only for a short amount of time early in the 
history of these systems. Relying on  Eq.~\ref{eq:PerGalaxyRate},  the formation rate of BH$^{\tzo}$s 
with mass of 2.5 to 5 $M_{\odot}$ is, 
\begin{eqnarray}
\Gamma_{\rm sc}^{BH^\tzo}(M^{\star}_{sc} , t)&=&(0.6-0.9) \times1.5 \times 10^{-4} \left(\frac{M^{\star}_{sc}}{10^{10.79} 
 M_{\odot}}\right) \nonumber \\
&\times& H(t-T_{1})H(T_{2}-t) \yr^{-1}.
\label{eq:PerStellarClusterRate}
\end{eqnarray}
$M^{\star}_{sc}$ is the mass in stars within a given stellar cluster, i.e. $M^{\star}_{sc} \sim 10^{5} \; M_{\odot}$ 
for a massive globular cluster. The source term at the globular clusters is taken for simplicity to be constant in time $t$ 
between $T_{1}$ and $T_{2}$, though the product of two Heaviside functions $H(t-T_{1})H(T_{2}-t)$. $T_{1}$ and $T_{2}$ are the timescales of collapse of a 18 $M_{\odot}$ (for $T_{1}$) and 
a 6 $M_{\odot}$ (for $T_{2}$) star. Approximately, $T_{1} = 10$ Myr and $T_{2} = 130$ Myr \footnote{Time 
of collapse or full loss of envelope is taken to be $\sim 1.2 \times$ the main sequence lifetime.}. 
We remind that our estimate relies on current star forming regions and may be different for 
the early stages of globular clusters.

\begin{figure}
    \centering
     \includegraphics[width=8.7cm,height=5cm]{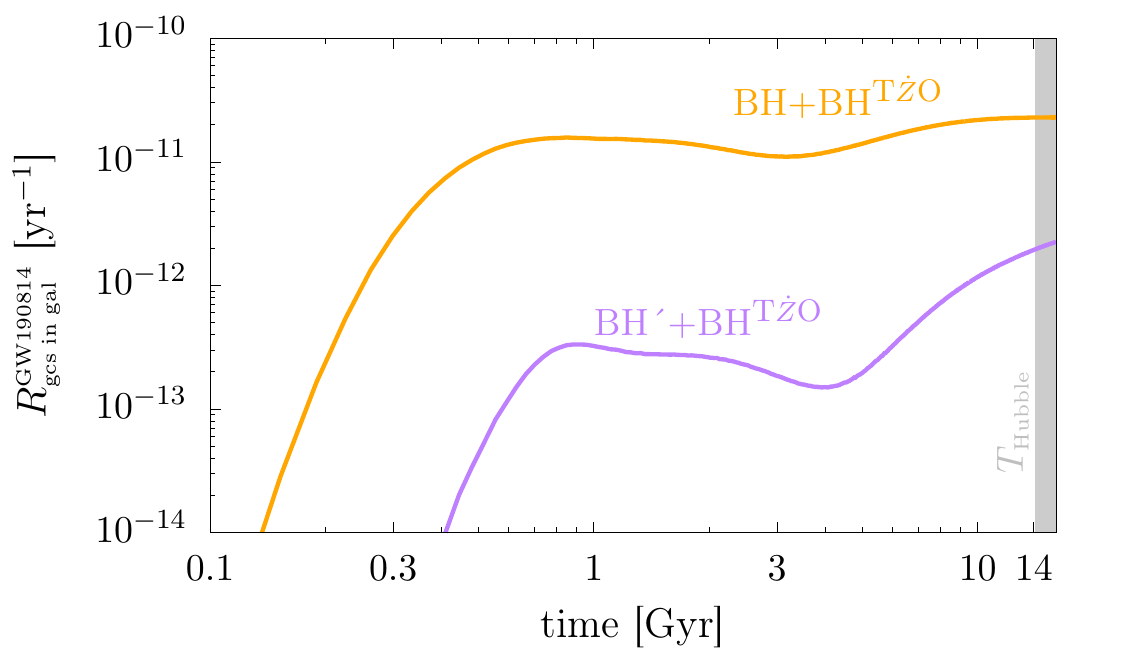} \\
     \caption{The merger rate of BH$^{\tzo}$ with 1st  generation black hole ``BH'' (orange) and 2nd generation 
     black holes ``\BHprime'' (purple) as it evolves in the Milky Way globular clusters. We are 
     assuming that all clusters were formed at the same time. }
    \label{fig:BHTZO_BH_mergers}
\end{figure}

The rate $\Gamma_{\rm sc}^{BH^\tzo}(M^{\star}_{sc})$ of Eq.~\ref{eq:PerStellarClusterRate}, can be used as a source 
term of low mass gap BHs inside clusters. In Ref.~\cite{Kritos:2021yty}, a numerical scheme was developed 
to model the dynamical interactions of low mass gap objects (as our BH$^{\tzo}$s) with regular mass stellar 
BHs,  massive second generation BHs and  stars. Once formed, the BH$^{\tzo}$s will first bind in binaries with 
stars. Those first binaries will then have exchange interactions with other BH$^{\tzo}$s  and more massive BHs, 
creating binaries composed solely by compact objects. At the same time since these new binaries are 
surrounded by stars, binary-single star interactions will take place resulting in the loose binaries breaking up 
and the hard ones becoming even tighter. In \cite{Kritos:2021yty}, all those interactions are included for a 
sequence of the observed globular cluster systems, taking into account their environmental parameters (densities 
and velocity distribution profiles). We implement the same code in this work taking all the observed Milky Way 
clusters as reference in order to evaluate a merger rate between low mass gap BH$^{\tzo}$ and BHs in the regular 
mass range. In Figure~\ref{fig:BHTZO_BH_mergers}, we show the total rate of mergers from all Milky Way 
globular clusters $\textrm{\textfrak{R}}_{\textrm{gcs in gal}}^{\rm GW190814}$.  From our results that rate can 
be taken to be roughly constant up to redshift of 1.
\begin{equation}
\textrm{\textfrak{R}}_{\textrm{gcs in gal}}^{\rm GW190814}(M^{\star}) = 2 \times 10^{-11} \left(\frac{M^{\star}}{10^{10.79} 
\, M_{\odot}}\right) \yr^{-1}. \; 
\label{eq:PerGalaxyRate_gw190814_from_GCs}
\end{equation}

Combining Eq.~\ref{eq:PerGalaxyRate_gw190814_from_GCs} with Eq.~\ref{eq:TZORateDensity} 
where in the place of $\textrm{\textfrak{R}}_{\rm gal}^{\tzo}$ we substitute with 
$\textrm{\textfrak{R}}_{\textrm{gcs in gal}}^{\rm GW190814}$, we get that the local GW190814-like 
event rate density of,
\begin{equation}
R^{\rm GW190814}_{\textrm{in gcs}}(z<0.1) = \small{\int_{M^{\star}_{\rm min}}^{M^{\star}_{\rm max}} d M^{\star} \; 
\textrm{\textfrak{R}}_{\textrm{gcs in gal}}^{\rm GW190814}(M^{\star}) \frac{\Phi(M^{\star})}{M^{\star}}}.
\label{eq:GW190814_RateDensity_from_GCs}
\end{equation}
This gives $R^{\rm GW190814}_{\textrm{in gcs}}(z<0.1) = 1.6 \times 10^{-4}$ $\Gpc^{-3} \yr^{-1}$. This 
result is about four orders of magnitude smaller than the reported GW190814 rate density. Thus,
globular clusters are excluded as the only environment where $\tzo$s can create a low mass gap black hole that 
subsequently via dynamical interactions, will form a binary with a BH and merge giving GW190814-type events. 

\textit{Forming binaries in nuclear star clusters:}
Nuclear star clusters come with a variety of masses between galaxies, however their stellar mass is related
 to the mass of the host galaxy \cite{2020ApJ...900...32P}, 
\begin{equation}
log_{10} (M_{\textrm{NSC}}) = 1.094 \cdot log_{10} \left( \frac{M^{\star}}{10^{6} M_{\odot}} \right) + 2.881.
\label{eq:NSCmass}
\end{equation}
Following \cite{2020ApJ...900...32P}, we can also evaluate the tidal radius $r_{t}$ of nuclear star clusters 
hosted by galaxies of given stellar mass. Taking all these clusters to have the same concentration parameter $c$ 
of 0.7, where $c = log_{10}(r_{t}/r_{c})$ and $r_{c}$ is the core radius where all BHs reside, and using the numerical 
code of \cite{Kritos:2021yty} we can evaluate the rate by which BH-BH$^{\tzo}$ merger events take place in
a nuclear star cluster hosted by a galaxy of given stellar mass. Properly weighting with the Schechter mass function for 
local galaxies in Eq.~\ref{eq:DoubleSchechtMF}, we get that the rate of mergers from all nuclear star 
clusters in galaxies $\textrm{\textfrak{R}}_{\textrm{nsc in gal}}^{\rm GW190814}$ is approximately described as, 
\begin{eqnarray}
\small{\textrm{\textfrak{R}}_{\textrm{nsc in gal}}^{GW190814}(M^{\star})} = \begin{cases} 10^{-15} \yr^{-1} \; \textrm{for} \;  M^{\star} < 10^{7} M_{\odot}, \\
10^{-15} \left(\frac{M^{\star}}{10^{7} \, M_{\odot}}\right) \yr^{-1} \; \textrm{for} \;  M^{\star} > 10^{7} M_{\odot}.
 \end{cases}
 \label{eq:NSCrate}
\end{eqnarray}
In turn that gives us a rate density from nuclear star clusters  that is $R^{\rm GW190814}_{\textrm{in nsc}}(z<0.1) 
= 1 \times 10^{-8}$ $\Gpc^{-3} \yr^{-1}$. 
Nuclear star clusters are very dense environments and as a result all BH-BH$^{\tzo}$ binaries even when formed 
will not remain binaries for long enough to merge via gravitational wave emission. Instead, in such dense environments
the merging BH binaries will contain approximately equal mass members. As an example, in a nuclear star cluster 
with size of $10^{8} M_{\odot}$, typically hosted by Milky Way sized galaxies, over a course of 10 Gyr there will be 
$\simeq 2 \times 10^{3}$ BH$^{\tzo}$ objects formed. Yet, the probability of even one of them merging with a stellar 
mass BH is estimated to be only $10^{-4}$. 

We note that nuclear star clusters do not have a clearly observed concentration-mass relation. 
Thus we have treated the concentration parameter $c$ as a free parameter.  The value of $c =0.7$ represents a 
low estimate of the allowed concentration in nuclear star cluster environments. That choice makes the BH-BH$^{\tzo}$ 
merger rates larger. Higher values of $c$ will only further suppress significantly the estimated rate of 
Eq.~\ref{eq:NSCrate}. 

Inside nuclear star clusters there is also the possibility of direct capture events between stellar mass BHs and BH$^{\tzo}$. 
Such events will create very hard binaries with high eccentricities that will rapidly merge via gravitational wave 
emission and may not undergo any exchange interactions \cite{OLeary:2008myb}. For the direct captures higher 
concentrations enhance the creation of tight BH-BH$^{\tzo}$  binaries. However, even for a very high value of 
concentration $c =2$, we get a rate density of $3 \times 10^{-6}$  $\Gpc^{-3} \yr^{-1}$ from direct capture events, 
that still is orders of magnitude smaller than the claimed rate for GW190814 type of events. 

\textit{Hierarchical triples in the field:}
As we described most $\tzo$s and in turn BH$^{\tzo}$s are created in the field with a rate density of  
$6\times 10^{2}$ $\Gpc^{-3} \yr^{-1}$. However, we need to include the probability that the BH$^{\tzo}$
will merge in a Hubble time with a stellar mass BH. Given the very low density of BHs in the field, the 
only possibility for BH-BH$^{\tzo}$  mergers in the field is that the final BH-BH$^{\tzo}$ binary originated from 
a hierarchical triple of a specific configuration. We denote $m_{1}$, $m_{2}$ and $m_{3}$ the Zero Age 
Main Sequence (ZAMS) star masses of the three initial stars, with $m_{1} > m_{2} > m_{3}$. The relevant 
configuration leading to a GW190814-like event has to be such that in the triple at its creation, the inner 
binary contained the stars of masses $m_{2}$ and $m_{3}$ in an orbit with semi-major axis $a_{\textrm{in}}$ 
and eccentricity $e_{\textrm{in}}$. The star with ZAMS mass $m_{2}$ has to be between 8 and up to 25 $M_{\odot}$, 
so that when it had its supernova (SN) explosion it created a NS. Instead, the least massive star with mass $m_{3}$ 
has to be between $6$ and $18 M_{\odot}$ giving us the red giant that with the NS will create the $\tzo$. 
Finally, $m_{1}$ has to be massive enough that it will
give a BH of mass 10-30 $M_{\odot}$, i.e. $m_{3}$ has to be at least  $ m_{3} > 30 M_{\odot}$. That star 
was on an initial outer orbit with respect to the $m_{2}-m_{3}$ binary with semi-major axis $a_{\textrm{out}}$ and 
eccentricity $e_{\textrm{out}}$.

The rate density of $6\times 10^{2}$ $\Gpc^{-3} \yr^{-1}$ for BH$^{\tzo}$s already accounts for the fact that 
the inner binary will survive the SN explosion of $m_{2}$ and that  $a_{\textrm{in}}$  is small enough for the 
$\tzo$ to form. To connect that rate density to a BH-BH$^{\tzo}$ merger rate density we need to include, first 
the fact that the $\tzo$ will be in a triple, second that the third object is the most massive member with a 
ZAMS mass of at least 30 $M_{\odot}$, third that the outer binary will survive both the SN kicks of $m_{1}$ and 
$m_{2}$, and fourth that once the BH-BH$^{\tzo}$ binary is created it will merge within a Hubble time. In the 
following we address the first three points. 

In \cite{2016ComAC...3....6T} it is noted that about 10$\%$ of low mass stars are in triples, with that fraction 
rising to about 50$\%$ for spectral type B stars.  Given the high value of the $m_{2}$ mass we take that 
about 30$\%$ of systems that give rise to $\tzo$s start out as triple systems.  Using the Kroupa initial stellar 
mass function we estimate that if the $m_{1}$ mass is independent of the masses of the other stars in the 
hierarchical triple system then only in 0.1-2 $\%$ of the triple systems, the $m_{1}$ mass will be $> 30 M_{\odot}$ 
\footnote{The $2\%$ fraction comes from taking into account the Kroupa mass function parametrization uncertainties 
such that more stars are predicted at its massive end.}. 
That is a very dominant suppression and at face value would bring the BH-BH$^{\tzo}$ merger rate density 
down to 0.2-4 $\Gpc^{-3} \yr^{-1}$ without including the third and fourth conditions. However, that suppression 
may be significantly mitigated by the fact that star systems likely form due to fragmentation processes. It is 
unlikely that the masses of the three stars are independent of each other. And since by having a $\tzo$ the 
mass $m_{2}$ is already massive enough, there may be an enhanced probability that the outer star is massive 
as well enhancing the  BH-BH$^{\tzo}$ merger rates. 

For these triples the mass ratio is $m_{1}/(m_{2} + m_{3}) \simeq 1$. Since we have already included the 
suppression factor of hierarchical triples, stability arguments give that the semi-major axes ratio of the 
outer orbit to the inner orbit  $a_{\textrm{out}}/a_{\textrm{in}}$ is between 5 and 20 \cite{2001MNRAS.321..398M}, 
a result that we also confirmed by semi-analytical calculations and is in agreement with numerical simulations 
and observations of triple systems \cite{2002A&A...384.1030S}. The initial (pre-SN explosions) 
eccentricity of the outer orbit of these triple systems follows a thermal distribution. Systems where the inner orbit 
will give a $\tzo$ are already very tight. Relying on observed orbital properties of binary systems \cite{1991A&A...248..485D}, 
we find that the equivalent triple system will not break up by the SN kicks 
of either the $m_{1}$ or $m_{2}$ stars as the typical SN kicks are only $\sim$100 km/s \cite{Mandel:2015eta}. In fact, 
the first natal kick of the $m_{1}$ SN explosion will marginally affect the system's binding energy decreasing 
the semi-major axis of the outer binary by a factor of $\simeq$3, which equals the fraction of the $m_{1}$ mass 
to that of its resulting BH remnant. 
We also find that the eccentricity distribution even after the SN explosion of $m_{1}$ is 
still going to be a thermal one. Thus, we find that effectively all triples where the inner binary will give a $\tzo$ 
survive the SN kicks of $m_{1}$ and $m_{2}$.  

Finally, we want to know the probability that the BH-BH$^{\tzo}$ will merge within a Hubble time. That binary has 
to be sufficiently tight. As noted in \cite{Zevin:2020gbd} in binary BH mergers from binary star systems, the objects
 get close to each other through a common envelope phase preceding the formation of the second black hole.  
 This would require a reliable estimate of the probability that the common envelope scenario takes place for our 
 case, which in turn requires specialized codes that take into account both stellar evolution and orbital dynamics 
 \cite{Zevin:2020gbd, 2016ComAC...3....6T}. Additional modifications to take into account the formation of $\tzo$ 
 and BH$^{\tzo}$ would be necessary to such codes. That is beyond the scope of the present work. If all BH-BH$^{\tzo}$ merge 
 within a Hubble time the merger rate density in the field from hierarchical triples can be 0.2-4 $\Gpc^{-3} \yr^{-1}$
 which is within the claimed LIGO-Virgo rate for GW190814-like events. Thus we find that it is important such 
 modifications to include $\tzo$s are performed in the future. Until then we are not yet able to rule this scenario out.

\begin{figure}
    \centering
    \includegraphics[width=8.7cm,height=5cm]{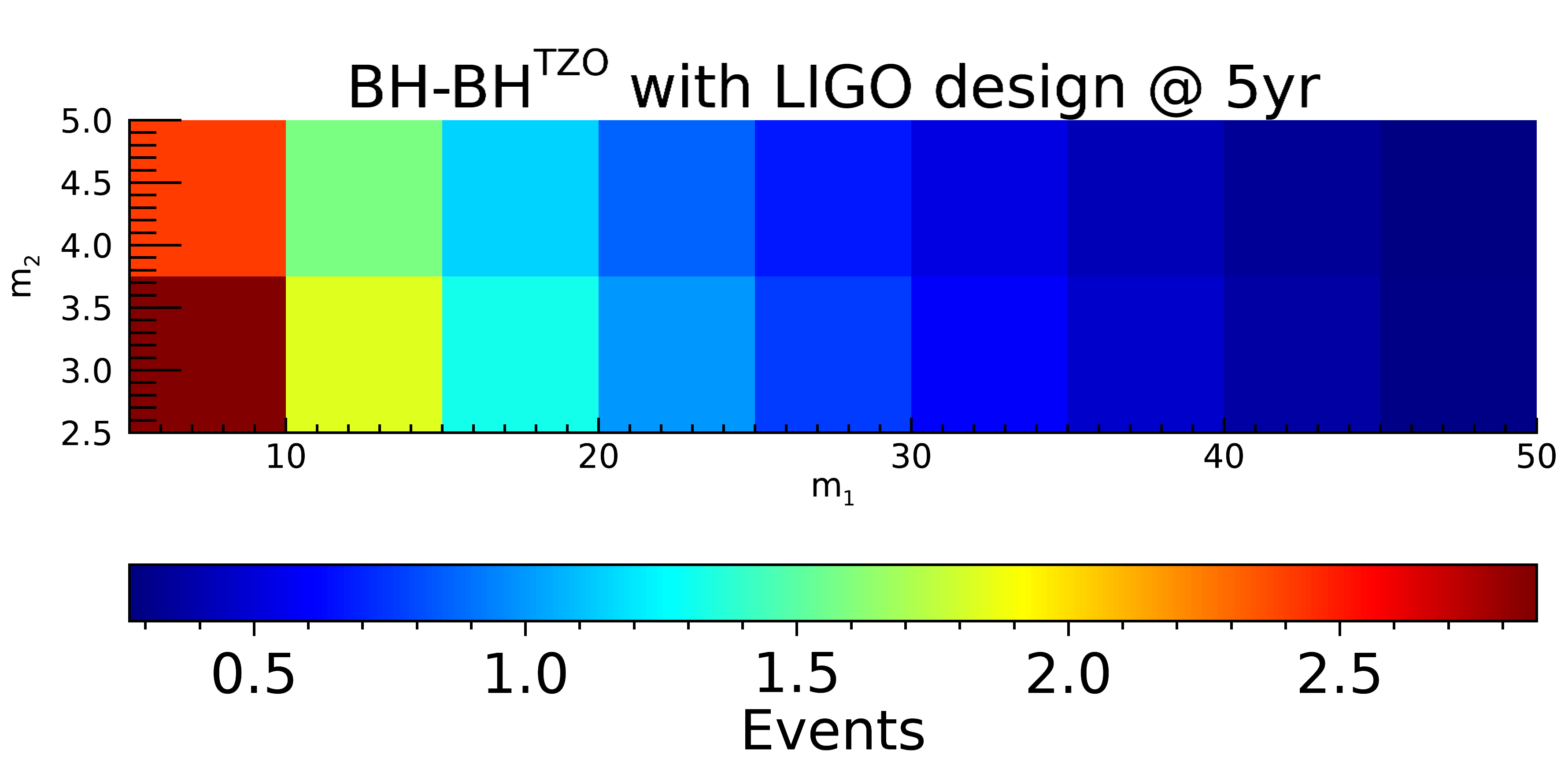} 
    \caption{Number of BH-BH$^{\tzo}$  events at S/N >8 at design sensitivity after 5 years.
   We take the rate density of these mergers to be $2 \Gpc^{-3} \yr^{-1}$.}
 \label{fig:Events}
\end{figure}

\textit{Observational perspectives:} 
As a final note, if the observed rate of  GW190814-like events is $\sim 2 \Gpc^{-3} \yr^{-1}$ then with the enhanced 
LIGO-Virgo-KAGRA sensitivity we expect a gradual filling up of the low mass gap range. The BH$^{\tzo}$s will cover 
the entire low mass gap range, as we show in Figure~\ref{fig:Events}. However, the most likely BH$^{\tzo}$s are the 
lower $\simeq 2.5 M_{\odot}$ as their mass comes from the NS accreting a fraction of the red giant's mass. That 
makes the observed $2.6 M_{\odot}$ mass of GW190814 a quite likely outcome.

\textit{Acknowledgements:} 
IC acknowledges support from the NASA Michigan Space Grant Consortium, Grant No. 80NSSC20M0124.
DG thanks the National Science Foundation for support in Grant No. PHY-1806219 and PHY-2102914. 

\bibliography{TZO_to_BBH}
\bibliographystyle{apsrev}

\end{document}